\documentclass[showpacs,amsmath,amssymb,twocolumn]{revtex4}
\usepackage{graphicx}
\begin{document}

\title{Meson Mixing in Pion Superfluid}
\author{\normalsize{Xuewen Hao and Pengfei Zhuang }}
\affiliation{Center of Theoretical Nuclear Physics, National
             Laboratory of Heavy Ion Collisions, Lanzhou 730000, China\\
             Physics Department, Tsinghua University, Beijing 100084, China}

\begin{abstract}
We investigate meson mixing and meson coupling constants in pion
superfluid in the framework of two flavor NJL model at finite
isospin density. The mixing strength develops fast with increasing
isospin chemical potential, and the coupling constants in normal
phase and in the pion superfluid phase behave very differently.
\end{abstract}


\maketitle

\section{Introduction}
\label{s1}
Recently, the study on the phase diagram of Quantum Chromodynamics
(QCD) is extended from finite temperature and baryon chemical
potential to finite isospin chemical potential. The physical
motivation to study pion superfluid formed at high isospin density
is related to the investigation of neutron stars, isospin asymmetric
nuclear matter and heavy ion collisions at intermediate energy.

Different approaches, such as chiral perturbation
theory\cite{son,kogut1,loewe,birse}, Nambu-Jona-Lasinio (NJL)
model\cite{toublan,frank,barducci1,he1,he2,he3,shao}, random matrix
method\cite{klein}, lattice QCD\cite{kogut2}, ladder
QCD\cite{barducci2} and strong coupling lattice QCD\cite{nishida},
have been used to investigate the QCD phase structure at finite
isospin density. It is found that there is a phase transition from
the normal phase to the pion superfluid at a critical isospin
chemical potential which is exactly the pion mass in the vacuum,
$\mu_I^c=m_\pi$. The temperature and chemical potential dependence
of meson masses in the pion superfluid is calculated in the NJL
model\cite{he1}, linear sigma model\cite{mao,sheikh} and $\phi^4$
theory\cite{jens}. The familiar meson properties are significantly
changed in the pion superfluid phase. In the normal phase which is
controlled by spontaneous chiral symmetry breaking and its
restoration, the scalar and pseudoscalar collective modes of a two
flavor quark system are $\sigma$ and $\pi$ mesons. However, they are
no longer the eigenmodes of the Hamiltonian in the pion superfluid,
and the new eigenmodes are the linear combination of them. It is the
combination that guarantees the Goldstone mode corresponding to the
spontaneous isospin symmetry breaking\cite{he1}.

To investigate the dynamic processes which may be considered as
signatures of the pion superfluid, one needs not only the meson
masses but also other in-medium meson properties in the superfluid.
In this Letter, we will study the hadron mixing and coupling
constants for $\sigma$ and $\pi$ mesons at finite isospin density.

It is well known that, the perturbation theory of QCD can well
describe the new phases at extremely high temperature and high
density, but the study on the phase structure at moderate
temperature and density depends on lattice QCD calculation and
effective models with QCD symmetries. One of the models that
enables us to see directly how the dynamic mechanisms of chiral
symmetry breaking and restoration operate is the NJL
model\cite{njl} applied to quarks\cite{njlquark}. When the
parameters in the model are properly chosen, the calculated chiral
phase transition line\cite{njlquark,zhuang} in temperature and
baryon chemical potential plane is very close to that obtained
with lattice QCD\cite{lattice}. Recently, this model is used to
investigate color superconductivity\cite{njlcsc} at moderate
baryon density and pion
superfluidity\cite{toublan,frank,barducci1,he1,he2,he3,shao} at
finite isospin density. We will study the meson mixing and
coupling constants in the pion superfluid in the framework of this
model.

The Letter is organized as follows. In Section \ref{s2} we review
the NJL model in mean field approximation for quarks and random
phase approximation (RPA) for mesons in the normal phase at finite
isospin chemical potential. In Section \ref{s3} we focus on the
meson mixing and meson couplings to quarks in the pion superfluid.
We summarize in Section \ref{s4}.

\section{Mesons in Normal Phase }
\label{s2}
The mesons in the normal phase at finite isospin chemical
potential are controlled by chiral dynamics and explicit isospin
symmetry breaking. In this section we review the quark propagator
in mean field approximation and meson polarizations in RPA in the
NJL model at finite temperature and baryon and isospin chemical
potentials. The Lagrangian density of the two flavor NJL model at
quark level is defined as\cite{njlquark}
\begin{equation}
\label{njl}
{\cal L} =
\bar{\psi}\left(i\gamma^{\mu}\partial_{\mu}-m_0+\mu\gamma_0\right)\psi
+G\Big[\left(\bar{\psi}\psi\right)^2+\left(\bar\psi
i\gamma_5\tau_i\psi\right)^2\Big]
\end{equation}
with scalar and pseudoscalar interactions corresponding to
$\sigma$ and $\pi$ excitations, where $m_0$ is the current quark
mass, $G$ is the four-quark coupling constant with dimension
(GeV)$^{-2}$, $\tau_i\ (i=1,2,3)$ are the Pauli matrices in flavor
space, and $\mu
=diag\left(\mu_u,\mu_d\right)=diag\left(\mu_B/3+\mu_I/2,\mu_B/3-\mu_I/2\right)$
is the quark chemical potential matrix with $\mu_u$ and $\mu_d$
being the $u$- and $d$-quark chemical potentials and $\mu_B$ and
$\mu_I$ being the baryon and isospin chemical potentials.

At zero isospin chemical potential, the Lagrangian density has the
symmetry of $U_B(1)\bigotimes SU_I(2)\bigotimes SU_A(2)$
corresponding to baryon number symmetry, isospin symmetry and
chiral symmetry. At finite isospin chemical potential, the isospin
symmetry $SU_I(2)$ and chiral symmetry $SU_A(2)$ are explicitly
broken to $U_I(1)$ and $U_A(1)$, respectively. Therefore, the
chiral symmetry restoration at finite isospin chemical potential
means only degeneracy of $\sigma$ and $\pi_0$ mesons, the charged
$\pi_+$ and $\pi_-$ behave still differently.

Introducing the chiral and pion condensates
\begin{eqnarray}
\label{condensate1} && \sigma = \langle\bar{\psi}\psi\rangle =
\sigma_u +\sigma_d,\nonumber\\
&& \sigma_u =\langle\bar u u\rangle,\ \ \ \sigma_d =\langle\bar d
d\rangle,\nonumber\\
&&\pi =\sqrt 2\langle\bar{\psi}i\gamma_5\tau_+\psi\rangle = \sqrt
2 \langle\bar{\psi}i\gamma_5\tau_-\psi\rangle
\end{eqnarray}
with $\tau_\pm =(\tau_1\pm i\tau_2)/\sqrt 2$ and taking all the
condensates to be real, the quark propagator in mean field
approximation can be expressed as a matrix in the flavor space
\begin{equation}
\label{quarkpropagator}
{\cal S}(p)= \left(\begin{array}{cc} {\cal S}_{uu}(p)&{\cal S}_{ud}(p)\\
{\cal S}_{du}(p)&{\cal S}_{dd}(p)\end{array}\right)
\end{equation}
with the four elements\cite{he1}
\begin{eqnarray}
\label{element}
{\cal S}_{uu} =
{\left(p_0+E+\mu_d\right)\Lambda_+\gamma_0\over
(p_0-E^-_-)(p_0+E^-_+)}+ {\left(p_0-E+\mu_d\right)\Lambda_-\gamma_0\over (p_0-E^+_-)(p_0+E^+_+)},\nonumber\\
{\cal S}_{dd} = {\left(p_0-E+\mu_u\right)\Lambda_-\gamma_0\over
(p_0-E^-_-)(p_0+E^-_+)}+{\left(p_0+E+\mu_u\right)\Lambda_+\gamma_0\over (p_0-E^+_-)(p_0+E^+_+)},\nonumber\\
{\cal S}_{ud} = {2iG\pi\Lambda_+\gamma_5\over
(p_0-E^-_-)(p_0+E^-_+)}+{2iG\pi\Lambda_-\gamma_5\over (p_0-E^+_-)(p_0+E^+_+)},\nonumber\\
{\cal S}_{du} = {2iG\pi\Lambda_-\gamma_5\over
(p_0-E^-_-)(p_0+E^-_+)}+{2iG\pi\Lambda_+\gamma_5\over
(p_0-E^+_-)(p_0+E^+_+)},
\end{eqnarray}
where $E^\pm_\mp = E^\pm \mp \mu_B/3$ are the energies of four
quasi-particles with $E^\pm = \sqrt{\left(E\pm
{\mu_I/2}\right)^2+4G^2\pi^2}$, $E = \sqrt{|{\bf p}|^2+M_q^2}$ and
effective quark mass $M_q=m_0-2G\sigma$, and $\Lambda_{\pm} =
\left[1\pm \gamma_0\left({\bf \gamma}\cdot{\bf
p}+M_q\right)/E\right]/2$ are the energy projectors. The quark
propagator (\ref{quarkpropagator}) is the background of our
following calculations for quarks and mesons. From the definitions
of the chiral and pion condensates (\ref{condensate1}), it is easy
to express them in terms of the matrix elements of the quark
propagator,
\begin{eqnarray}
\label{gap1}
&&\sigma_u = - \sum_p Tr\left[i {\cal
S}_{uu}(p)\right],\nonumber\\
&& \sigma_d = - \sum_p Tr\left[i {\cal
S}_{dd}(p)\right],\nonumber\\
&& \pi = \sum_p Tr \left[\left({\cal S}_{ud}(p)+{\cal
S}_{du}(p)\right)\gamma_5\right],
\end{eqnarray}
where the trace $Tr=Tr_C Tr_D$ is taken in color and Dirac spaces,
and the four momentum integration is defined as
$\sum_p=iT\sum_n\int d^3{\bf p}/(2\pi)^3$ with
$p_0=i\omega_n=i(2n+1)\pi T\ (n=0,\pm1,\pm2,\cdots)$ at finite
temperature $T$. Note that, for $\mu_B=0$ or $\mu_I=0$ there is
always $\sigma_u=\sigma_d$, since in this case the chemical
potential difference between $\bar u $ and $u$ is the same as that
between $\bar d$ and $d$. The coupled set of gap equations
(\ref{gap1}) determines self-consistently the three condensates.
The isospin chemical potential dependence of the effective quark
mass $M_q$ and pion condensate $\pi$ is explicitly shown in
\cite{he1}. The phase transition from the normal phase to the pion
superfluid is of second order.

In the NJL model, the meson modes are regarded as quantum
fluctuations above the mean field. The two quark scattering via
meson exchange can be effectively expressed at quark level in
terms of quark bubble summation in RPA\cite{njlquark}. The quark
bubbles, namely the meson polarization functions are defined
as\cite{zhuang}
\begin{equation}
\label{polarization1}
\Pi_{mn}(k) = i\sum_p Tr \left(\Gamma_m^*
{\cal S}(p+k)\Gamma_{n} {\cal S}(p)\right)
\end{equation}
with the trace $Tr = Tr_C Tr_F Tr_D$ taken in color, flavor and
Dirac spaces and the meson vertexes
\begin{equation}
\label{vertex} \Gamma_m = \left\{\begin{array}{ll}
1 & m=\sigma\\
i\tau_+\gamma_5 & m=\pi_+ \\
i\tau_-\gamma_5 & m=\pi_- \\
i\tau_3\gamma_5 & m=\pi_0\ ,
\end{array}\right.\ \
\Gamma_m^* = \left\{\begin{array}{ll}
1 & m=\sigma\\
i\tau_-\gamma_5 & m=\pi_+ \\
i\tau_+\gamma_5 & m=\pi_- \\
i\tau_3\gamma_5 & m=\pi_0\ . \\
\end{array}\right.
\end{equation}
The explicit $T, \mu_B$ and $\mu_I$ dependence of the meson
polarization functions (\ref{polarization1}) used in the following
discussion can be found in Appendix B of \cite{he1}.

At $\mu_I\le\mu_I^c$ where $\mu_I^c$ is the critical isospin
chemical potential for the pion condensate, the system is in the
normal phase with diagonal quark propagator, and the bubble
summation in the effective interaction in RPA selects its specific
isospin channel by choosing at each stage the same proper
polarization\cite{njlquark}. Therefore, the meson masses $M_m
(m=\sigma,\pi_+,\pi_-,\pi_0)$ which are determined by poles of the
meson propagators at $k_0=M_m$ and ${\bf k}=0$ are related only to
their own polarization functions\cite{njlquark},
\begin{equation}
\label{mass1}
1-2G\Pi_{mm}(k_0)|_{k_0=M_m} = 0.
\end{equation}
From the comparison of these mass equations with the gap equation
for the pion condensate $\pi$ at $T=\mu_B=0$ but finite $\mu_I$,
the critical isospin chemical potential where the normal phase
ends and the pion superfluid starts is exactly the pion mass in
the vacuum\cite{kogut2,he1,he2}, $\mu_I^c=m_\pi$, and the $\mu_I$
dependence of the meson masses is simple\cite{he1}, $
M_\sigma(\mu_I) = m_\sigma,\ M_{\pi_0}(\mu_I)= m_\pi$ and
$M_{\pi_\pm}(\mu_I)= m_\pi\mp\mu_I$, where $m_\sigma$ and $m_\pi$
are the $\sigma$ and $\pi$ masses at $T=\mu_B=\mu_I=0$. The
isospin neutral mesons keep their vacuum masses, while the isospin
charged mesons change their masses linearly. These relations hold
until the pion condensate starts.

The meson couplings to quarks, $g_{mq\bar q}$, are related to the
residues at the corresponding poles of the meson
propagators\cite{njlquark},
\begin{equation}
\label{gmqq1}
g_{mq\bar q}^2 =
\left[\partial\Pi_{mm}(k_0)/\partial k_0^2\right]^{-1}|_{k_0=M_m}.
\end{equation}

\section {Mesons in Pion Superfluid}
\label{s3}
In the phase with finite pion condensate which leads to spontaneous
isospin symmetry breaking, the quark propagator contains
off-diagonal elements, we must consider all possible isospin
channels in the bubble summation in RPA\cite{njlquark}. While there
is no mixing between $\pi_0$ and other mesons\cite{he1},
$\Pi_{\pi_0\sigma}(k) = \Pi_{\pi_0\pi_+}(k) = \Pi_{\pi_0\pi_-}(k) =
0$, the other three mesons are coupled together, and the effective
interaction in two quark scattering via exchanging these mesons in
RPA becomes a summation in the meson space,
\begin{equation}
\label{ueff1} U(k)=\Gamma_m^*{\cal M}_{mn}(k) \Gamma_n,\ \ \
m,n=\sigma,\ \pi_+,\ \pi_-
\end{equation}
with the meson matrix ${\cal M}(k)$ defined by
\begin{equation}
\label{calm1}
{\cal M}(k)={2G\over 1-2G\Pi(k)}={2G\over
D(k)}\overline{\cal M}(k),
\end{equation}
where $1-2G\Pi(k)$ is the meson polarization matrix\cite{he1}
\begin{eqnarray}
\label{polarization3}
&& 1-2G\Pi \\
 =&&
\left(\begin{array}{ccc}
1-2G\Pi_{\sigma\sigma}&-2G\Pi_{\sigma\pi_+}
&-2G\Pi_{\sigma\pi_-}\\
-2G\Pi_{\pi_+\sigma}&1-2G\Pi_{\pi_+\pi_+}&-2G\Pi_{\pi_+\pi_-}\\
-2G\Pi_{\pi_-\sigma}&-2G\Pi_{\pi_-\pi_+}&1-2G\Pi_{\pi_-\pi_-}\\
\end{array}\right),\nonumber
\end{eqnarray}
$D(k)$ is its determinant,
\begin{equation}
\label{d} D(k)=\det\left(1-2G\Pi(k)\right),
\end{equation}
and $\overline{\cal M}(k)$ is defined as $\overline{\cal
M}(k)=D(k)/(1-2G\Pi(k))$. In this case, $\sigma, \pi_+$ and
$\pi_-$ are no longer the eigenmodes of the Hamiltonian, the new
eigenmodes are linear combinations of them. In the following we
call these new eigenmodes in the pion superfluid phase as
$\overline\sigma, \overline\pi_+$ and $\overline\pi_-$.

The $\pi_0$ mass and coupling constant are still controlled by its
own polarization function at ${\bf k}=0$,
\begin{eqnarray}
\label{pi0}
&& 1-2G\Pi_{\pi_0\pi_0}(k_0)|_{k_0=M_{\pi_0}}=0,\nonumber\\
&& g_{\pi_0q\bar q}^2 =
\left[\partial\Pi_{\pi_0\pi_0}(k_0)/\partial
k_0^2\right]^{-1}|_{k_0=M_{\pi_0}},
\end{eqnarray}
since it is independent of the other mesons. At $T=\mu_B=0$, the
$\pi_0$ mass is exactly equal to the isospin chemical
potential\cite{he1}, $M_{\pi_0}(\mu_I)=\mu_I$. At the critical
point $\mu_I^c=m_\pi$, it is continuous with the solution
$M_{\pi_0}=m_\pi$ in the normal phase.

The masses of the new eigenmodes $\overline\sigma, \overline\pi_+$
and $\overline\pi_-$ are determined by the pole of the effective
interaction at ${\bf k}=0$,
\begin{equation}
\label{mass4}
D(k_0)|_{k_0=M_\theta}=0,\ \ \
\theta=\overline\sigma,\overline\pi_+,\overline\pi_- .
\end{equation}
It can be proven analytically that there is always a zero solution
which guarantees the Goldstone mode, $M_{\overline\pi_+}=0$,
corresponding to the spontaneous isospin symmetry
breaking\cite{he1}.

In order to derive the coupling constants for the new modes, we
first expand the effective interaction $U$ around the meson mass
$k_0^2=M_\theta^2$ at ${\bf k}=0$,
\begin{equation}
\label{ueff2}
U(k_0)\simeq {2G\over
(dD(k_0)/dk_0^2)|_{k_0=M_\theta}}{\Gamma_m^*\overline{\cal
M}_{mn}(M_\theta)\Gamma_n\over k_0^2-M_\theta^2},
\end{equation}
and then make a transformation from the coupled meson space
($\sigma, \pi_+, \pi_-$) to the diagonalized meson space
($\overline\sigma, \overline\pi_+, \overline\pi_-$). With the help
of the pole equation (\ref{mass4}) for the mass $M_\theta$, we can
derive the relations between the diagonal and off-diagonal elements
of the matrix $\overline{\cal M}$,
\begin{eqnarray}
\label{mmmprf}
&& \overline{\cal M}_{mm} \overline{\cal M}_{nn} -
\overline{\cal M}_{mn}^2 = (1 - 2G \Pi_{ll}) D = 0,\nonumber\\
&& l,m,n=\sigma, \pi_+, \pi_-,\ \ \ l\ne m \ne n,
\end{eqnarray}
where all the quantities are evaluated at $k_0=M_\theta$. Taking
into account the symmetric property of $\overline{\cal M}$,
\begin{equation}
\label{symmetry}
\overline{\cal M}_{mn}(M_\theta)=\overline{\cal
M}_{nm}(M_\theta),
\end{equation}
the effective interaction $U$ via changing a $\theta$-meson in the
diagonalized meson space can be written as
\begin{equation}
\label{ueff3}
U(k_0)\simeq {2G \overline
M(M_\theta)\over(dD(k_0)/dk_0^2)|_{k_0=M_\theta}}{\Gamma_\theta^*\Gamma_\theta\over
k_0^2-M_\theta^2}
\end{equation}
in terms of the new meson vertex
\begin{eqnarray}
\label{nbt}
&& \Gamma_\theta =\sum_m \sqrt{\overline{\cal
M}_{mm}(M_\theta)}\ \Gamma_m \Big/\sqrt{\overline
M(M_\theta)},\nonumber\\
&& \Gamma_\theta^* =\sum_m \sqrt{\overline{\cal
M}_{mm}(M_\theta)}\ \Gamma_m^* \Big/\sqrt{\overline M(M_\theta)},
\end{eqnarray}
where $\overline M(M_\theta)$ is defined as $\overline
M(M_\theta)=\sum_m\overline {\cal M}_{mm}(M_\theta)$.

Since the meson coupling constant is defined as the residue of the
effective interaction at the pole, we can extract the
$\theta$-meson coupling constant $g_{\theta q\bar q}$ from
(\ref{ueff3}),
\begin{equation}
\label{gmqq2}
g_{\theta q \bar q}^2 = {2G \overline
M(M_\theta)\over(dD(k_0)/dk_0^2)|_{k_0=M_\theta}}.
\end{equation}

To explicitly describe the meson mixing in the pion superfluid
phase, we can introduce mixing angles between two mesons, for
instance, the angels $\alpha$ between $\pi_+$ and $\sigma$,
$\beta$ between $\pi_-$ and $\sigma$ and $\gamma$ between $\pi_+$
and $\pi_-$ in the $\overline\sigma$-meson channel,
\begin{eqnarray}
\label{angels}
\tan\alpha &=& \sqrt{\overline{\cal
M}_{\pi_+\pi_+}(M_{\overline\sigma})}\Big/\sqrt{\overline{\cal
M}_{\sigma\sigma}(M_{\overline\sigma})},\nonumber\\
\tan\beta &=& \sqrt{\overline{\cal
M}_{\pi_-\pi_-}(M_{\overline\sigma})}\Big/\sqrt{\overline{\cal
M}_{\sigma\sigma}(M_{\overline\sigma})},\nonumber\\
\tan\gamma &=& \sqrt{\overline{\cal
M}_{\pi_+\pi_+}(M_{\overline\sigma})}\Big/\sqrt{\overline{\cal
M}_{\pi_-\pi_-}(M_{\overline\sigma})}\nonumber\\
&=&\tan\alpha/\tan\beta.
\end{eqnarray}
It is easy to see that only $\alpha$ and $\beta$ are independent.
The mixing angels in the $\overline\pi_+$ and $\overline\pi_-$
channels can be defined in a similar way.

Since the NJL model is non-renormalizable, we can employ a hard
three momentum cutoff $\Lambda$ to regularize the gap equations for
quarks and pole equations for mesons. In the following numerical
calculations, we take the current quark mass $m_0=5$ MeV, the
coupling constant $G=4.93$ GeV$^{-2}$ and the cutoff $\Lambda=653$
MeV\cite{zhuang}. This group of parameters corresponds to the pion
mass $m_\pi=134$ MeV, the pion decay constant $f_\pi=93$ MeV and the
effective quark mass $M_q=310$ MeV in the vacuum.

\begin{figure}
\centering
\includegraphics[width=8cm]{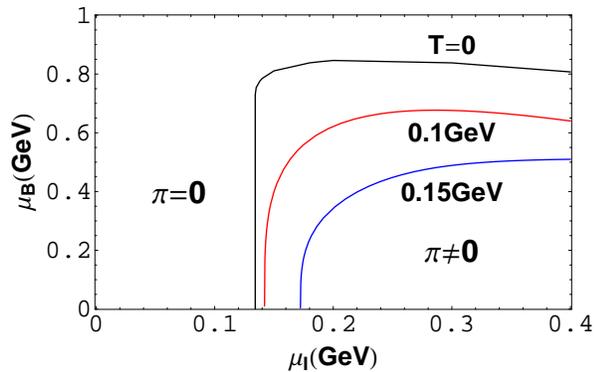}
\caption{The phase diagram in $\mu_I-\mu_B$ plane at fixed
temperature $T$. } \label{fig1}
\end{figure}

The phase diagram in the $\mu_I-\mu_B$ plane at fixed temperature
$T$ is shown in Fig.\ref{fig1}. The system is in the normal phase
at low $\mu_I$ or high $\mu_B$ and in the pion superfluid at high
$\mu_I$ and low $\mu_B$. At zero temperature, the critical isospin
chemical potential $\mu_I^c$ does not change with $\mu_B$ till
\begin{equation}
\mu_B=3(M_q-m_\pi/2)\sim 730\ \text{MeV}
\end{equation}
which is determined by the comparison of the gap equation for the
pion condensate at $\pi=0$ and $T=0$ with the pion mass equation in
the vacuum. With increasing temperature, the critical value
$\mu_I^c$ increases and the pion superfluid region decreases, due to
the melting of pion condensate in hot mediums.

\begin{figure}
\centering
\includegraphics[width=8cm]{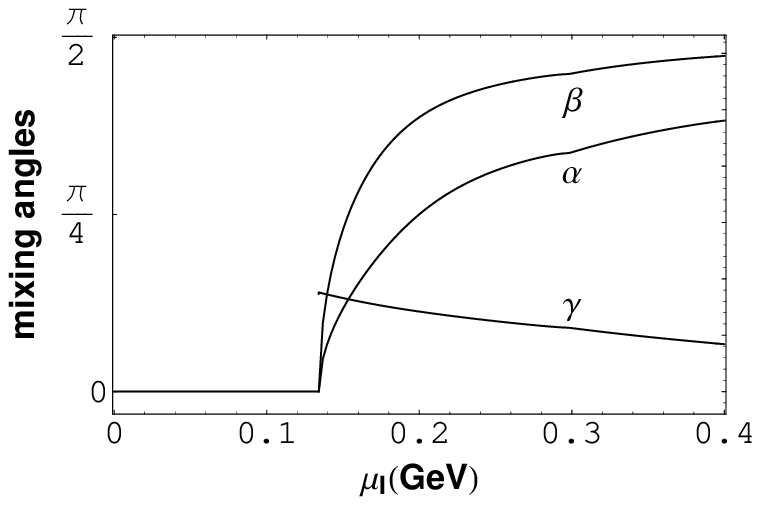}
\includegraphics[width=8cm]{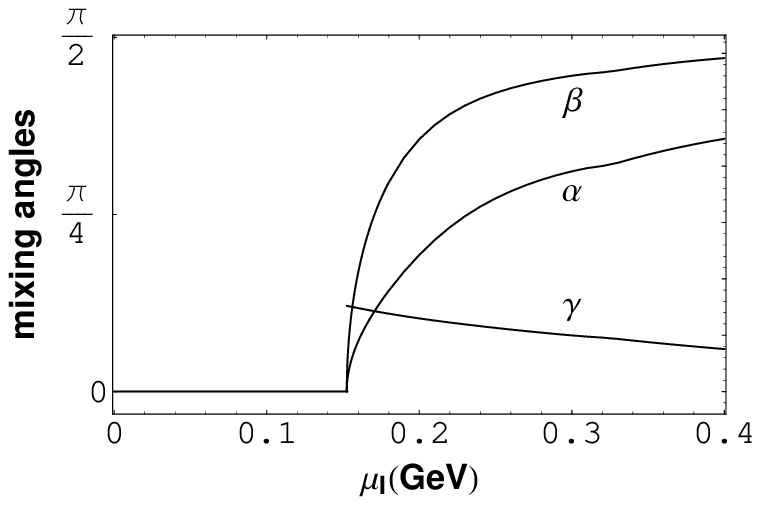}
\caption{The mixing angles $\alpha, \beta$ and $\gamma$ in the
$\overline\sigma$-meson channel as functions of $\mu_I$ at
$T=\mu_B=0$ (upper panel) and $T=0.1$GeV and $\mu_B=0.4$GeV (lower
panel).}
\label{fig2}
\end{figure}

The mixing angles $\alpha, \beta$ and $\gamma$ in the
$\overline\sigma$-meson channel are shown in the upper panel of
Fig.\ref{fig2} as functions of $\mu_I$ at $T=\mu_B=0$. In the
normal phase with $\mu_I\le \mu_I^c$, $\sigma$ and $\pi$
themselves are the collective excitation modes of the system, and
there is no mixing among them. In the pion superfluid phase with
$\mu_I>\mu_I^c$, $\alpha$ and $\beta$ indicate the $\pi_+ -
\sigma$ and $\pi_- - \sigma$ mixing strengths in the
$\overline\sigma$-meson channel, and $\gamma$ reflects the
relative strength between them. While in the very beginning of the
superfluid any mixing is weak, it develops fast. For $\mu_I>200$
MeV which corresponds to $\alpha=\pi/4$, the mixing is already so
strong that the contribution from $\pi_+$ to $\overline\sigma$ is
larger than that from $\sigma$. Similarly, the $\pi_-$-component
in $\overline\sigma$ becomes more important than the
$\sigma$-component itself even for $\mu_I>150$ MeV. Therefore, at
not very high isospin chemical potential the $\pi_-$- and
$\pi_+$-components start to dominate the $\overline\sigma$ mesons.
While the $\pi_- - \sigma$ mixing is always stronger than the
$\pi_+ - \sigma$ mixing, namely $\beta > \alpha$, the relative
strength shown by $\gamma$ decreases with increasing isospin
density.

At zero temperature, since the phase structure of the pion
superfluid does not change till the baryon chemical potential
$\mu_B=730$ MeV, see the vertical straight line in Fig.\ref{fig1},
the mixing angles at finite baryon chemical potential $\mu_B<730$
MeV are exactly the same as the ones shown in the upper panel of
Fig.\ref{fig2}. Since the pion superfluid happens only at low
temperature, we calculated the $\mu_I$ dependence of the meson
mixing angels at $T=0.1$ GeV and $\mu_B=0.4$ GeV, shown in the lower
panel of Fig.\ref{fig2}. From the comparison with the upper panel,
the shape of the mixing angles at finite temperature and baryon
chemical potential is almost the same as that at $T=\mu_B=0$, the
only remarkable change is the critical value $\mu_I^c$. It increases
from $134$ MeV at $T=\mu_B=0$ to $153$ MeV at $T=0.1$ GeV and
$\mu_B=0.4$ GeV.

From our numerical calculation, the mixing angles in the
$\overline\pi_-$-meson channel behave similarly. However, the case
is significantly changed for the Goldstone mode $\overline\pi_+$.
The vertex for $\overline\pi_+$ can be greatly simplified as
\begin{equation}
\label{vertexpi+}
\Gamma_{\overline\pi_+} = \left(\Gamma_{\pi_+} -
\Gamma_{\pi_-}\right)/\sqrt 2
\end{equation}
at any temperature and baryon and isospin chemical potentials. The
Goldstone mode contains only $\pi_+$- and $\pi_-$-components, and
the fractions for the two components are exactly the same.

\begin{figure}
\centering
\includegraphics[width=8cm]{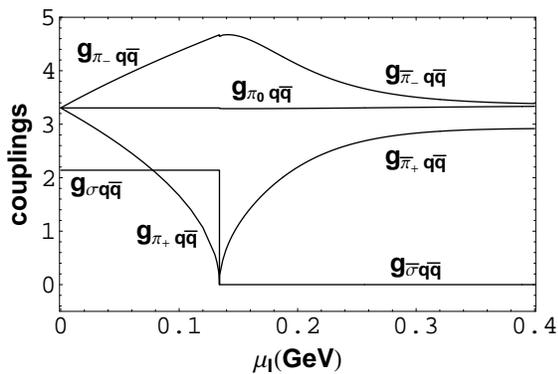}
\caption{The coupling constants for $\sigma, \pi_0, \pi_+, \pi_-$
in the normal phase and $\overline\sigma, \pi_0, \overline\pi_+,
\overline\pi_-$ in the pion superfluid phase as functions of
$\mu_I$ at $T=\mu_B=0$.} \label{fig3}
\end{figure}

The meson couplings to quarks are shown in Fig.\ref{fig3} as
functions of $\mu_I$ at $T=\mu_B=0$. In the normal phase with
$\mu_I<\mu_I^c$, the coupling constants are calculated through the
diagonal polarization functions $\Pi_{mm}$. At $T=\mu_B=0$,
$\Pi_{mm}$ depends only on the quark mass $M_q$, meson mass $M_m$
and isospin chemical potential $\mu_I$. Since $M_m$ for isospin
neutral mesons $\sigma$ and $\pi_0$ and $M_q$ are constants, the
coupling constants $g_{\sigma q\bar q}$ and $g_{\pi_0 q \bar q}$ are
$\mu_I$ independent. In the pion superfluid phase, the coupling
constants, determined by the diagonal and off-diagonal polarization
functions $\Pi_{mn}$, behave very differently. From the mass
relation $M_{\overline\pi_-}\to M_{\pi_0}$ at $\mu_I\to \infty$, the
couplings $g_{\overline\pi_- q \bar q}$ and $g_{\pi_0 q \bar q}$
approach each other at large enough $\mu_I $. Note that $g_{\pi_0 q
\bar q}$ is no longer a constant but changes slowly in the pion
superfluid. Since we did not consider the meson widths in the pole
equation (\ref{mass4}), the condition for a meson to decay into a
$q$ and a $\bar q$ is that its mass lies above the $q-\bar q$
threshold. The meson masses, calculated through the pole equations
(\ref{mass1}) in the normal phase and (\ref{pi0}) and (\ref{mass4})
in the pion superfluid, are shown in Fig.15 of \cite{he1}. In the
pion superfluid, the heaviest mode is $\overline\sigma$ and its mass
is beyond the threshold value. As a result, there exists no
$\overline\sigma$ meson, and the coupling constant
$g_{\overline\sigma q \bar q}$ drops down to zero at the critical
point $\mu_I^c$ and keeps zero at $\mu_I>\mu_I^c$. For the other
three mesons $\pi_0, \overline\pi_+$ and $\overline\pi_-$, their
masses are below the threshold value, and the coupling constants are
nonzero.

From Fig.\ref{fig2}, the meson mixing angels are only slightly
changed by finite temperature $T$ and baryon chemical potential
$\mu_B$, when $T$ and $\mu_B$ are in a reasonable region. Therefore,
it can be expected that the meson coupling constants will not be
significantly changed by finite $T$ and $\mu_B$, except a remarkable
shift of the critical isospin chemical potential $\mu_I^c$ from
$134$ MeV to a higher value.

\section{Summary}
\label{s4}
We have investigated the meson mixing and meson coupling constants
in the NJL model at finite isospin chemical potential. In the pion
superfluid phase, the normal mesons are no longer the collective
excitation modes of the system, and the mixing among them becomes
important. For the Goldstone mode, it contains only charged pions
and their fractions are exactly the same in the whole superfluid
region. For the other new eigenmodes, the meson mixing starts to
control the system at $\mu_I\gtrsim 150$ MeV which is only a
little bit higher than the critical value $\mu_I^c=m_\pi=134$ MeV
at $T=\mu=0$. The coupling constants for the conventional mesons
in the normal phase and for the new eigenmodes in the pion
superfluid phase behave very differently. The splitting of meson
mass and coupling constant due to explicit and spontaneous isospin
symmetry breaking can be used to further calculate the $\pi\pi$
scattering and the number ratio $\pi_+/\pi_-$ in isospin
asymmetric matter, which can help us to measure the meson
properties at finite isospin density.

\noindent {\bf \underline{Acknowledgments:}} The work is supported
by the grants NSFC10428510, 10435080, 10575058 and
SRFDP20040003103.


\begin{thebibliography}{99}
\bibitem{son}      D.T. Son, M.A. Stephanov, Phys. Rev. Lett. {\bf 86}, 592(2001); Phys. Atom. Nucl. {\bf 64}, 834(2001).
\bibitem{kogut1}   J.B. Kogut, D. Toublan, Phys. Rev. {\bf D64}, 034007(2001).
\bibitem{loewe}    M. Loewe, C. Villavicencio, Phys. Rev. {\bf D67}, 074034(2003); Phys. Rev. {\bf D}70, 074005(2004).
\bibitem{birse}    M.C. Birse, T.D. Cohen, J.A. McGovern, Phys. Lett. {\bf B516} 27(2001).
\bibitem{toublan}  D. Toublan, J.B. Kogut, Phys. Lett. {\bf B564} 212(2003).
\bibitem{frank}    M. Frank, M. Buballa, M. Oertel, Phys. Lett. {\bf B562}, 221(2003).
\bibitem{barducci1}A. Barducci, R. Casalbuoni, G. Pettini, L. Ravagli, Phys. Rev. {\bf D69}, 096004(2004).
\bibitem{he1}      L. He, M. Jin, P. Zhuang, Phys. Rev. {\bf D71}, 116001(2005).
\bibitem{he2}      L. He, P. Zhuang, Phys. Lett. {\bf B615}, 93(2005).
\bibitem{he3}      L. He, M. Jin, P. Zhuang, Phys. Rev. {\bf D74}, 036005 (2006).
\bibitem{shao}     G. Shao, L. Chang, Y. Liu, X. Wang, Phys. Rev. {\bf D73}, 076003(2006).
\bibitem{klein}    B. Klein, D. Toublan, J.J.M. Verbaarschot, Phys. Rev. {\bf D68}, 014009(2003).
\bibitem{kogut2}   J.B. Kogut, D.K. Sinclair, Phys. Rev. {\bf D66}, 034505(2002).
\bibitem{barducci2}A. Barducci, G. Pettini, L. Ravagli, R. Casalbuoni, Phys. Lett. {\bf B564} 217(2003).
\bibitem{nishida}  Y. Nishida, Phys. Rev. {\bf D69}, 094501(2004).
\bibitem{mao}      H. Mao, N. Petropoulos, S. Shu, W. Zhao, J. Phys. {\bf G32}, 2187(2006); N. Petropoulos, hep-ph/0402136.
\bibitem{sheikh}   M.L. El-Sheikh, M. Loewe, hep-ph/0701100.
\bibitem{jens}     Jens O. Andersen, Phys. Rev. {\bf D75}, 065011(2007).
\bibitem{njl}      Y. Nambu, G. Jona-Lasinio, Phys. Rev. {\bf 122}, 345(1961); Phys. Rev. {\bf 124}, 246(1961).
\bibitem{njlquark} See, for reviews, U. Vogl, W. Weise, Prog. Part. Nucl. Phys. {\bf 27}, 195(1991); S.P. Klevansky, Rev.
                   Mod. Phys. {\bf 64}, 649(1992); M.K. Volkov, Phys. Part. Nucl.
                   {\bf 24}, 35(1993); T. Hatsuda, T. Kunihiro, Phys. Rept. {\bf 247},
                   338(1994); M. Buballa, Phys. Rept. {\bf 407}, 205(2005).
\bibitem{zhuang}   P. Zhuang, J. H\"ufner, S.P. Klevansky, Nucl. Phys. {\bf A576}, 525(1994).
\bibitem{lattice}  See, for instance, S. Ejiri et. al., Prog. Theor. Phys. Suppl. {\bf 153},
                   118(2004); Z. Fodor, S. D. Katz, Prog. Theor. Phys. Suppl. {\bf 153}, 86(2004).
\bibitem{njlcsc}   T.M. Schwarz, S.P. Klevansky, G. Papp, Phys. Rev. {\bf C60}, 055205(1999);
                   M. Huang, P. Zhuang, W. Chao, Phys. Rev.{\bf D65}, 076012(2002), Phys. Rev.{\bf D67}, 065015(2003);
                   I. Shovkovy, M. Huang, Phys. Lett. {\bf B564}, 205(2003).
\end{thebibliography}
\end{document}